# Upper critical field and irreversibility line in superconducting $MgB_2$

G. Fuchs[*], K.-H. Müller, A. Handstein, K. Nenkov[1], V.N. Narozhnyi[2], D. Eckert, M. Wolf, L. Schultz

*Institut für Festkörperforschung und Werkstoffforschung Dresden, Postfach 270116, D-01171 Dresden, Germany*

**Abstract.**

The upper critical field $H_{c2}(T)$ of sintered pellets of the recently discovered $MgB_2$ superconductor was investigated by transport, *ac* susceptibility and *dc* magnetization measurements in magnetic fields up to 16 T covering a temperature range between $T_c \sim 39$ K and $T = 3$ K $\sim 0.1 T_c$. The temperature dependence of the upper critical field, $H_{c2}(T)$, shows a positive curvature near $T_c$ similar to that found for the borocarbides $YNi_2B_2C$ and $LuNi_2B_2C$ indicating that $MgB_2$ is in the clean limit. The irreversibility line was consistently determined from *dc* magnetization measurements and from the imaginary component of *ac* susceptibility. The irreversibility field was found to increase up to 8.5 T at 10 K.

## 1. Introduction

The recent discovery of superconductivity in $MgB_2$ [1] at temperatures as high as 40 K has stimulated considerable interest in this system. $MgB_2$ which has a hexagonal $AlB_2$ structure is a type II-superconductor. A significant boron isotope effect was observed [2] which is an indication for electron-phonon mediated superconductivity in this compound. Superconducting parameters as the Ginsburg-Landau parameter $\kappa = 26$ [3] and the temperature dependence of the upper critical field $H_{c2}(T)$ [3-6] were determined from transport and magnetization measurements [3-7]. So far, a nearly complete $H_{c2}(T)$ curve was reported for a $MgB_2$ wire sample showing a high residual resistivity ratio of about 25 [7]. In the present investigation, the temperature dependence of the upper critical field of a sintered $MgB_2$ pellet was studied in magnetic fields up to 16 T in order to analyse the shape of $H_{c2}(T)$ in a wide temperature range for a sample with a moderate residual resistance ratio.

## 2. Experimental

Polycrystalline samples of $MgB_2$ were prepared by a conventional solid state reaction. A stoichiometric mixture of Mg and B was pressed into pellets. These pellets were wrapped in a Ta foil and sealed in a quartz ampoule. The samples were sintered at 950°C for two hours. Electrical resistance and the superconducting transition of a sample 5 mm in length with a cross-section of about 1 mm$^2$ (cut from the initially prepared pellet) were investigated in magnetic fields up to 16 T using the standard four probe method and current densities between 0.2 and 1 A/cm$^2$. *Ac* susceptibility and *dc* magnetization measurements were performed on other pieces from the same pellet in magnetic fields up to 9 T and 5 T, respectively.

---

[*] Corresponding author. Tel.: +49 351 4659 538; fax: +49 351 4659 537
  *E-mail address*: fuchs@ifw-dresden.de
[1] On leave from: Int Lab. of High Magn. Fields, Wroclaw; ISSP-BAS, Sofia, Bulgaria
[2] On leave from Inst. for High Pressure Physics, Russian Acad. Sci., Troitsk, Russia

## 3. Results and discussion

In Fig. 1, the temperature dependence of the electrical resistance of the investigated sample is shown. The resistivity at 40 K and 300 K are about 6.4 $\mu\Omega$cm and 29 $\mu\Omega$cm, respectively, resulting in a residual resistance ratio (RRR) of approximately 4.5. The midpoint value of the superconducting transition at zero-magnetic field is 38.8 K. A similar $T_c$ value of $T_c = 39.0$ K was determined from $ac$ susceptibility data using the onset temperature of the superconducting transition. The field dependence of the electrical resistance of the same sample is shown in Fig. 2 for several temperatures between 36 and 2.9 K. A considerable broadening of the transition curves is observed at low temperatures which is caused at least partially by flux-flow effects at high magnetic fields. The transition widths gradually broaden from 0.2 T at 36 K to 5 T at 6 K. In Fig. 3 the field values $H_{10}$, $H_{50}$ and $H_{90}$ defined at 10%, 50% and 90% of the normal-state resistance are plotted as functions of temperature. Identical results have been found from resistance-vs.-temperature transition curves measured at different magnetic fields [8]. Additionally, Fig. 3 shows upper critical field data determined from $dc$ magnetization and from $ac$ susceptibility measurements. The onset of superconductivity was used to define $H_{c2}$ from $ac$ susceptibility. An example for the determination of $H_{c2}$ from $dc$ magnetization is shown in Fig. 4, where magnetization data are plotted for $T = 27.5$ K in an expanded view. The large resolution allows to visualize not only $H_{c2}$, but also the irreversibility field $H_{irr}$ and a large region between $H_{c2}$ and $H_{irr}$ in which the change of magnetization is reversible. The comparison of the upper critical fields obtained from $dc$ magnetization, $ac$ susceptibility and resistance measurements in Fig. 3 shows clearly, that for the investigated sample $H_{c2}^{mag}$ ($H_{c2}$ from magnetization) coincides with $H_{90}$, whereas $H_{c2}^{sus}$ ($H_{c2}$ from susceptibility) agrees approximately with $H_{10}$. This behaviour is unusual because typically the onset of superconductivity determined from $ac$ susceptibility agrees well with the midpoint value ($H_{50}$) of the normal-state resistivity. The difference between $H_{c2}^{mag}$ and $H_{c2}^{sus}$ can be explained by sample inhomogeneities. It seems that the major part of the sample has the reduced upper critical fields measured by $ac$ susceptibility, whereas only a relatively small fraction of the sample shows higher $H_{c2}$ values. One has to take into account that already a narrow current path through the sample with improved parameters is sufficient to produce the observed resistive-transition data. The properties of this small fraction can be detected and systematically investigated by sensitive magnetization measurements (see Fig. 5). The extrapolation of $H_{10}(T)$ to $T = 0$ yields an upper critical field of $H_{c2}(0) \sim 13$ T for the major fraction of the sample, whereas for the small fraction with improved parameters, $H_{c2}(0) \sim 18$ T is estimated by extrapolation of $H_{90}(T)$ to $T = 0$. Using these $H_{c2}(0)$ values, the coherence lengths $\xi_o = [\phi_o/(2\pi H_{c2}[0])]^{0.5}$ are found to be 5.0 nm (major fraction) and 4.2 nm (small fraction).

A peculiarity of the $H_{c2}(T)$ dependence shown in Fig. 3 is its pronounced positive curvature near $T_c$. Such a positive curvature of $H_{c2}(T)$ near $T_c$ is a typical feature observed for the non-magnetic rare-earth nickel borocarbides $R$Ni$_2$B$_2$C ($R$=Y, Lu) and can be explained by taking into account the dispersion of the Fermi velocity using an effective two-band model for superconductors in the clean limit [9]. This model can be successfully applied to MgB$_2$, as was shown very recently [10]. We conclude that also our MgB$_2$ samples are within the clean limit in spite of the rather moderate RRR value of 4.5.

The $H_{c2}(T)$ curves in Fig 3 can be described, in a wide temperature range $0.3T_c < T < T_c$, by the simple expression

$$H_{c2} = H_{c2}^*(1-T/T_c)^{1+\alpha}, \qquad (1)$$

where $H_{c2}^*$ and $\alpha$ are fitting parameters. It should be noted that $H_{c2}^*$ differs from the true value of $H_{c2}(0)$ due to the negative curvature of the $H_{c2}$-vs.-$T$ curve observed at low temperatures. The fit curves in Fig. 3 describing the $H_{10}(T)$ and $H_{90}(T)$ data between 12 K and $T_c$ correspond to values of $\alpha = 0.25$ and $\alpha = 0.32$, respectively. Similar values for the parameter $\alpha$ describing the positive curvature of $H_{c2}(T)$ are known from the rare-earth nickel borocarbides $YNi_2B_2C$ and $LuNi_2B_2C$ [11].

It is interesting to note that the $H_{c2}(T)$ curve reported for a high quality $MgB_2$ wire with a RRR value of about 25 shows an almost linear temperature dependence in an extended temperature range. In this case, a positive curvature was observed only in a narrow temperature range near $T_c$. It is also remarkable, that the width of the resistive superconducting transition of this wire at low temperatures is similar to that of our sintered sample. In particular at $T = 1.5$ K, onset and completion of superconductivity (corresponding in our notation approximately to $H_{90}$ and $H_{10}$, respectively) were reported to be at 16.2 T and 13 T, respectively.

The magnetization-vs-field curves shown in Fig. 4 are reversible between the upper critical field and the irreversibility field $H_{irr}$. The so-called irreversibility line $H_{irr}(T)$ of a superconductor separates reversible and irreversible regions in the field-temperature phase diagram and is well established for all high-$T_c$ superconductors. The knowledge of the irreversibility line is important for the most potential applications because non-zero critical currents are confined to magnetic fields below this line. The irreversibility field of the investigated $MgB_2$ sample was determined (i) from $M$-vs.-$H$ curves using the onset of irreversibility as shown in Figs. 4 and 5 and (ii) from the peak value of the imaginary part of the $ac$ susceptibility measured with an $ac$ field amplitude of 0.1 Oe at a frequency of 133 Hz. Both methods yield comparable results as illustrated in Fig. 3. The irreversibility line shown in Fig. 3 lies well below the $H_{c2}(T)$ curve of the major fraction of the sample, however, the irreversibility field at low temperatures is (with $H_{irr}(10\ K) \sim 8.5$ T) higher than for other $MgB_2$ samples for which $H_{irr}(10\ K) \sim 6$ T was reported [3,12]. The observed shift of the irreversibility line to higher magnetic fields in the investigated sample seems to suggest improved pinning of flux lines, however, one has to take into account that the irreversibility field determined from $M$-vs.-$H$ curves is influenced by the sweep rate $dH_a/dt$ of the applied magnetic field $H_a$. The reason is that the value of $\Delta M \propto j_c^*$ (and the corresponding critical current density $j_c^*$) at which the irreversibility field is defined, strongly depends on the level $E_c \propto dH_a/dt$ of the electrical field induced on the sample surface. The data shown in Fig. 3 correspond to a sweep rate of $dH_a/dt = 40$ Oe/sec resulting in $E_c = 0.04$ µV/cm and $j_c^* = 0.2$ A/cm$^2$ provided that the current circulates mainly over the entire sample which has a diameter of about 2 mm. Lower irreversibility fields would be obtained by using lower sweep rates or by defining $H_{irr}$ at a higher level of $j_c^*$. For instance, relatively low irreversibility fields were obtained by using a $j_c^*$ criterion of 30 A/cm$^2$ to determine $H_{irr}$ [12].

It is interesting to compare the results for the investigated $MgB_2$ sample with $RRR = 4.5$ with the data reported for a polycrystalline $MgB_2$ sample with $RRR \approx 20$ [3] in the field range up to 9T. Whereas the $H_{c2}$ data at temperatures $T > 20$ K are similar in both cases, higher $H_{c2}$ values are found for the sample with $RRR = 4.5$ at lower temperatures. In particular, the value of $H_{10}=10.4$ T obtained for this sample at $T = 10$ K is approximately by 1 Tesla higher than the corresponding $H_{c2}$ at 10 K of the sample with $RRR \approx 20$. At the same time the improved pinning in the sample investigated in this paper results in enhanced irreversibility fields compared with those found for the sample with the higher $RRR$ value [3].

Noteworthy is the pronounced positive magnetoresistance MR of the investigated sample in the normal state reaching ~20% in an applied field of $H = 16$ T at $T = 40$ K, see Fig. 2. Although this value is smaller than reported for a polycrystalline sample with RRR ≈ 20 (~80% at $H = 9$ T [3]), the magnetoresistance is clearly visible in contrast to results for $MgB_2$ samples prepared under high pressure [13]. In these samples with RRR~2.5 the normal state resistivity was found to be independent of the applied magnetic field. It is interesting to note that large values of the normal state MR were observed earlier for high-quality nonmagnetic polycrystalline $LuNi_2B_2C$ samples [14]. The reason for the large MR in $MgB_2$, which is obviously varying with sample quality characterized by the value of RRR is not clarified so far.

## 4. Conclusion

In conclusion, the upper critical field of $MgB_2$ was investigated in a wide temperature range between 3 K and $T_c$. The onset of the superconducting transition of *ac* susceptibility measurements was found to agree with $H_{c2}$ data measured resistively at 10% of the normal state resistance and represents the upper critical field of the major fraction of the investigated sample. A small fraction of the sample exhibits higher upper critical fields, which were measured both resistively and by sensitive *dc* magnetization measurements. The investigated sample shows a variation of the extrapolated upper critical field at $T = 0$ between $H_{c2}(0) = 13$ T and about 18 T. A significant positive curvature observed for $H_{c2}(T)$ in a wide temperature region $0.3T_c < T \leq T_c$ suggests that the investigated $MgB_2$ sample is within the clean limit. The irreversibility line $H_{irr}(T)$ determined from *dc* magnetization and *ac* susceptibility measurements was found to increase up to $H_{irr} = 8.5$ T at 10 K. However, it is important to note that systematic investigations are necessary in order to separate the influence of pinning properties and measurement conditions on the irreversibility line of $MgB_2$.

**Acknowledgement**

The authors thank S.V. Shulga and S.-L. Drechsler for valuable discussions.

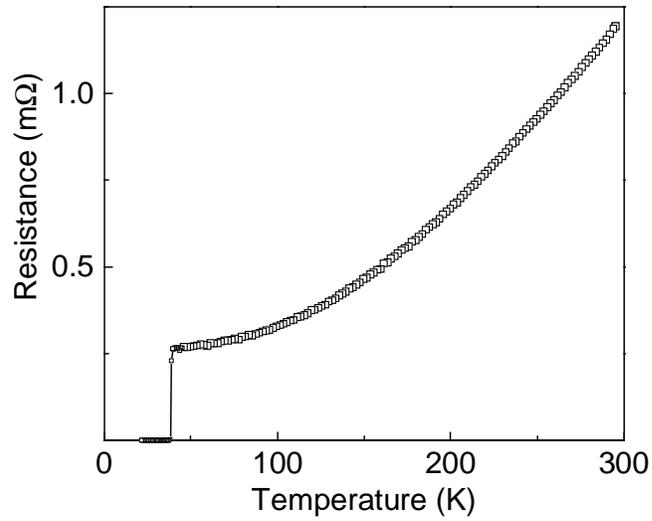

**Fig. 1.** Temperature dependence of the resistance of a polycrystalline $MgB_2$ sample in zero applied field.

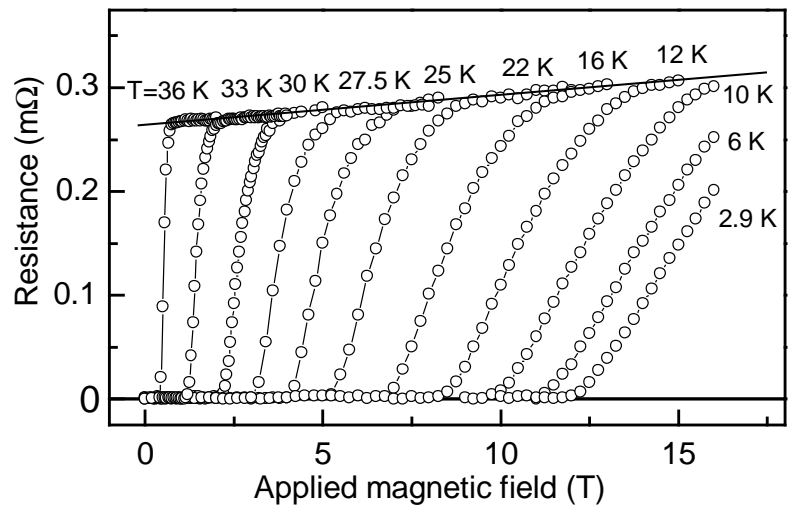

**Fig. 2.** Field dependence of the resistance of the $MgB_2$ sample of Fig. 1 for several temperatures between 36 and 2.9 K.

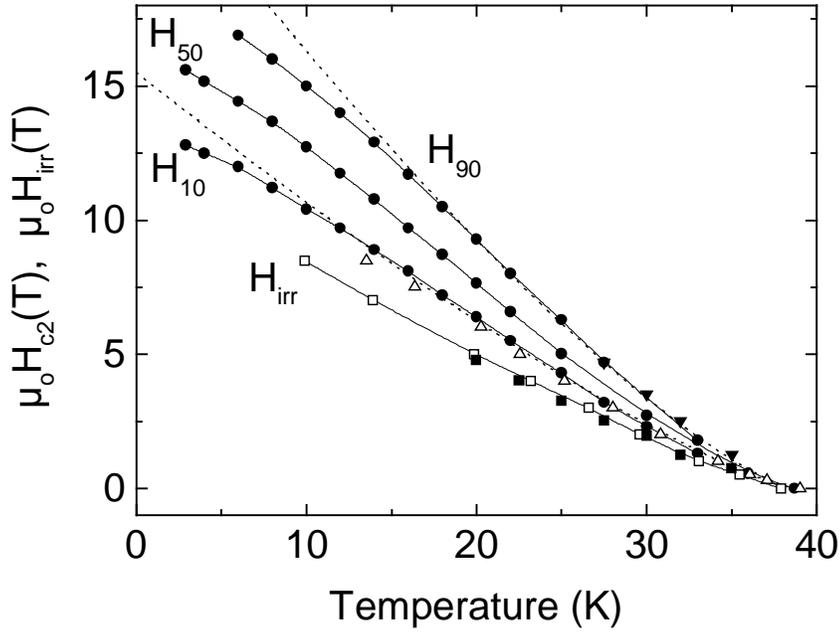

**Fig. 3.** Temperature dependence of the upper critical field determined from *dc* magnetization (▼), *ac* susceptibility (△) and resistivity measurements (●) defined at 10% ($H_{10}$), 50% ($H_{50}$) and 90% ($H_{90}$) of the normal state resistance and of the irreversibility field determined from *dc* magnetization (■) and from *ac* susceptibility (□). Dotted lines are calculated $H_{c2}(T)$ curves using Eqn. (1) by fitting $H_{c2}^*$ and $\alpha$ to the experimental data. Solid lines are guides for eye. Sample as in Fig. 1.

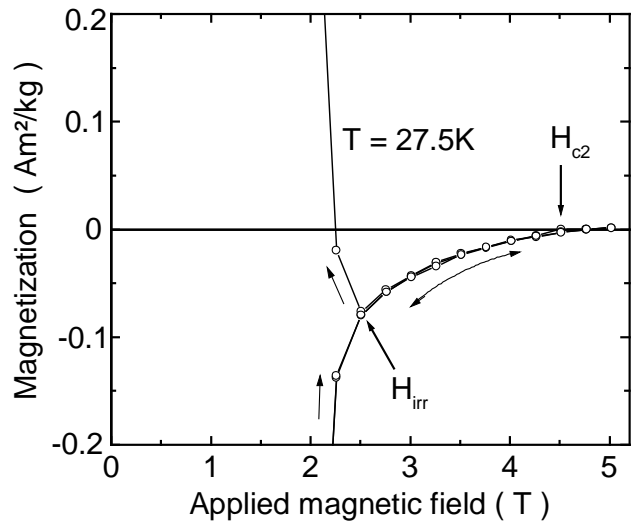

**Fig. 4.** Expanded view of *dc* magnetization vs. applied field at $T = 27.5$ K of a $MgB_2$ sample taken from the same pellet as that of Fig. 1. Arrows mark the upper critical field $H_{c2}$ and the irreversibility field $H_{irr}$. Arrows are also used to show how the applied field was changed in the different branches of the magnetization loop.

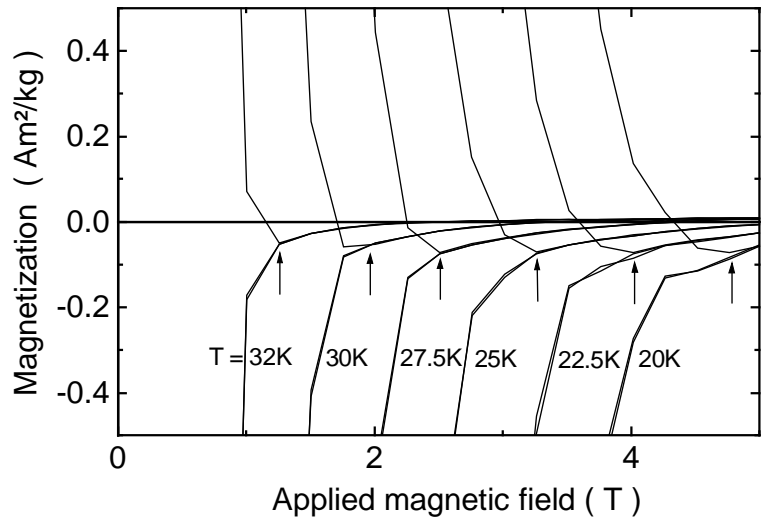

**Fig. 5.** Magnetization vs. applied field for several temperatures between 20 K and 35 K. The irreversibility fields $H_{irr}$ are marked by arrows. Sample as in Fig. 4.